\documentclass[fleqn,usenatbib]{mnras}

\usepackage{newtxtext,newtxmath}
\usepackage{anyfontsize}
\usepackage{graphicx}

\usepackage{placeins}

\usepackage[T1]{fontenc}

\DeclareRobustCommand{\VAN}[3]{#2}
\let\VANthebibliography\thebibliography
\def\thebibliography{\DeclareRobustCommand{\VAN}[3]{##3}\VANthebibliography}

\usepackage{graphicx}	
\usepackage{amsmath}	
\usepackage{indentfirst}

\usepackage{float}

\usepackage{orcidlink}

\title{TIC 393818343 c: Discovery and characterization of a Neptune-like planet in the Delphinus constellation}

\author[G. Conzo et al.]{
G. Conzo$^{\orcidlink{0000-0002-2412-1558}}$,$^{\hyperlink{i1}1}$\thanks{E-mail: giuconzo@gmail.com}
N. Leiner$^{\orcidlink{0009-0008-8978-4441}}$,$^{\hyperlink{i2}2}$
K. Lynch,
M. Moriconi$^{\orcidlink{0000-0002-5600-0953}}$,$^{\hyperlink{i1}1}$
N. Ruocco$^{\orcidlink{0009-0006-0421-8957}}$,$^{\hyperlink{i3}3}$
and T. Scarmato$^{\orcidlink{0000-0003-2463-6268}}$ $^{\hyperlink{i4}4}$
\hypertarget{i1}\\
{$^{1}$}Gruppo Astrofili Palidoro, Fiumicino, Italy
\hypertarget{i2}\\
{$^{2}$}Johnston Community College, Smithfield, North Carolina, US
\hypertarget{i3}\\
{$^{3}$}Osservatorio Astronomico Nastro Verde MPC Code C82, Sorrento, Italy
\hypertarget{i4}\\
{$^{4}$}Toni Scarmato’s Observatory MPC Code L92, San Costantino, Briatico, Italy\\
}

\pubyear{\the\year{}}

\begin{document}
\label{firstpage}
\pagerange{\pageref{firstpage}--\pageref{lastpage}}
\maketitle

\begin{abstract}
We report on the statistical confirmation of a second planet inside the TIC 393818343 system. The first planet \textit{TIC 393818343 b} has been confirmed and classified as a Warm Jupiter planet with a period of $P = (16.24921 \pm 0.00003)$ days. The second planet in the system has an orbital period of $P = (7.8458 \pm 0.0023)$ days and orbits 2.05 times closer to its host star. The second planet was initially spotted by the Las Cumbres Observatory (LCOGT) and amateur astronomers. This Super-Neptunian exoplanet marks TIC 393818343 as a multi-planetary system.
\end{abstract}




\section{Introduction}

While studying the Transit Timing Variations for the first planet, the researchers noticed discrepancies in the predicted orbital period of TIC 393818343 b (\cite{2024RNAAS...8...53C} and \cite{2024AJ....168...26S}). The offset in transit times was more than one hour which led to the suspicion and confirmation of TIC 393818343 c. It is classified as a Super-Neptunian exoplanet with a radius of $Rp = (0.078019 \pm 0.000006)R_{j}$, a Transit Duration of $Tdur \sim (3.71 \pm 0.0045) hours$ and a reported orbital period of $P= (7.8458 \pm 0.0023) days$, Super-Neptunian planets are relatively rare around stars like our Sun. This is likely due to the fact of the stars age, mass, and it's metallicity (\cite{2023AJ....166..209M}).

Data from the Transiting Exoplanet Survey Satellite (TESS) (\cite{2015JATIS...1a4003R}) was utilized, but the team found no evidence of the suspected second planet. This was due from both the dilution factor and TESS's large pixel size of 21 arcseconds per pixel. TIC 393818343 has a reported dilution of approximately $1.03\%$ due to starlight leakage from the foreground star, TIC 393818340.

Limb darkening is an optical effect resulting from the blackbody radiation of the host star, which can cause sharp variations in flux ($\Delta f$) due to the emitted radiation. As a result, further chromaticity studies were warranted to better understand these variations in the context of this study. Limb darkening parameters were estimated via ExoFast (\cite{2013PASP..125...83E}). Misinterpretation of the light curves can lead to false exoplanet types and radius estimates (\cite{2012CQGra..29x5002C}). Coincidentally, sometimes they can be classified as eclipsing binaries due to the sharp variations in most cases of astrophysical false positive scenarios (\cite{2011MNRAS.418.1165H}). 

Since radial velocity was not able to be conducted, synthetic observations were obtained and had a relative mass determination from a sigma confidence level of $\sigma =0.707$ of this planet being Neptunian in size. We used the \textbf{Forecaster} package following our manual radius estimates derived from the transit depth overtime. This Python package takes epoch and flux values with uncertainties and calculates probabilities ranging from Stellar, Terrestrial, Neptunian and Jovian in percentage rates (\cite{2017ApJ...834...17C}).
While this method is unorthodox, it gives us a rough estimate of what to expect in radial velocity calculations from ground-based observatories. However, the downfall for this method, is that it does not take into account stellar noise, instrumental noise and long term instability of gravitational influences should more planets be presented in analysis (\cite{2016ApJ...829L...9V}). Another factor in synthetic radial velocity is that actual radial velocity calculations account for stellar rotation, which broadens the spectral lines. By measuring the wobble through Doppler shift as the planet completes one orbit, astronomers can derive the planet’s orbital eccentricity.

\section{Photometric observations}

\subsection{Exoplanet Transit and Differential Photometry}

The use of differential photometry, which is applied to the transit of extra-solar planets, has shown that amateur instruments can reach a proven level of precision with their telescopes. HD 17156 b (\cite{2007A&A...476L..13B}), whose brightness variation is only a magnitude 0.005 marks the extreme challenges for the use of differential photometry. 
	
In the process of differential photometry, it is necessary to collect as many photon as possible from the star that the planet is orbiting, and the coinciding reference stars to then compare the resulting figures. Photons that are gathered from a specific source depends crucially on the brightness. The smaller depths (for example 1/10,000 ppm) is a depth that you would find for a terrestrial sized planet. This would have to be observed over a longer period of time, and would require more extensive follow ups as they could easily mistaken for astrophysical false positives.

This basic formula (Equation \ref{eq:ADU}) calculates the relationship between the flux of the star in question and the total flux of the other reference stars. The technique of differential photometry uses comparison stars. Comparison stars are nearby stars that have the same magnitude or the same level of \textit{background Analog Digital Units (ADUs)}. The relative change in the brightness of any variation could be caused by atmospheric conditions or instrumental effects which ultimately complicate the investigation. This is mitigated by warranted follow ups by applying different filters to affect the chromaticity of the object this in term allows the team to observe any sharp variations in delta flux (\cite{2024arXiv240409143K}). 

Limb darkening parameters are estimated from the stellar blackbody radiation emitted from the star to account for the sharp variation in flux, it also helps with stellar variability and most astrophysical false positives. Comparison stars are then selected which measure the change in brightness over a series of hours instead of days, since this is ground-based photometry and not space-based. In the case of TIC 393818343 c, we estimate a relative radius from the transit depth of a Super-Neptunian sized exoplanet with a radius of $Rp = (0.078019 \pm 0.000006)R_{j}$ across which supports the Super-Neptunian estimates. Using the equation below, it gives us the radius of the planet if three consecutive transits are seen with the same variations.

\begin{equation}
    Flux=\frac{AduStar}{{\sum} Adu(ref_{1},ref_{2},ref_{3}...)}.
	\label{eq:ADU}
\end{equation}

This formula gives a relationship between the flow of the star that is under investigation and the total flow of the stars of reference. To normalize the values obtained, the following equation is used (Equation \ref{eq:FLUX})

\begin{equation}
    \frac{FLUX(Star)}{Median}.
	\label{eq:FLUX}
\end{equation}

Using differential photometry, one can then construct a graph using a linear function that provides the best fit for the data obtained. If the star is less than a magnitude of 11, it has been shown that the blurring of the telescope can increase the accuracy of the data, if the range of the photon flow is sufficiently narrowed, which should be between $70\%$ and $90\%$ of the CCD dynamic range. The blur method is a method that allows you to collect the same number of photons but spread over an area greater than the CCD camera. Using this method, it became clear that the homogeneous flow and the shrinking of data scintillation increased the error on the measured ADU. The brighter a star is, the more focus on the star is required in order to obtain more precise data (\cite{2014arXiv1409.2693S}). 

\subsection{Methodology}

The methodology being used is to compare the flux of the object under investigation with the flux of objects that are in the same field. Shooting the images, like we said earlier, should preferably be done with an \textit{R Johnson Cousin filter} to obtain photometric data. The image series should be as long as possible, at least two hours, and have the correct exposure times. This is to have the highest possible signal-to-noise ratio \textit{(S/N)} while taking into account that the exposure times should not be too long, to not have overexposed images. If the telescope is on an equatorial mount, it must be well aligned with the pole to have minimized problems with bending and tracing. We stress that the object can be tracked so it can have the appropriate data retrieved. It is important to include a field of view range that is large enough to contain the flow, but not too large to incorporate any background stars that might have significant dilution near the source (\cite{2018EPSC...12...66M}). If it is possible, you must maintain stable temperature, which should only deviate by around $0.5^{\circ}C$ during the time spent gathering data. 

If you are unsure about which calibration file is the most important, it would be darks. Darks allow you to clarify the images which in term clarifies the data. Shooting darks in succession would obtain a master dark. A master dark is a file that shows all possible temperatures that occurred throughout the recovery, which will allow the pictures to be clearer and allow precise data to be pulled from the object. 
Depending on the brightness of the object, the quantum efficiency of the camera, and the \textit{R filter} that you use, allows you to reduce the total quantity of photons arriving in the CCD compared to shooting without a filter. This is unless you have a formula to get the appropriate time optimum exposure. If the counting of the ADU is equal to $70\%$ of the level of saturation of the camera, then you can proceed to capture the full range of images. If you want a better $S/N$ it must be taken into account that the ADU should not exceed $90\%$ of the level of saturation (\cite{2018haex.bookE.117D}).

Once the images are finalized, we must then have to calibrate them with then with a program that can measure the ADU of individual items automatically so that you get the count for each image (\cite{2009astro2010S.155K}). The program then measures the individual sources in the individual images, providing an output file with dates in JD, ADU, and the values of the objects for each image records in chronological order (Table \ref{tab:example_photometry}).

\begin{table}
	\centering
	\caption{Sample records with the measured data. In the first column is the Julian Date for each image, in subsequent columns are the counts of the objects that you want photometry of the star in question and the reference stars.}
	\label{tab:example_photometry}
	\begin{tabular}{lccr} 
		\hline
		JD & FLUX & Ref1 & Ref2\\
		\hline
		2454718.3129500 & 21304 & 510508 & 44780\\
		2454718.3136991 & 22548 & 521433 & 45027\\
		2454718.3144502 & 22145 & 518231 & 39568\\
		\hline
	\end{tabular}
\end{table}

Since the program can only do automatic photometry on 5 items at a time, if you have more than 5 photometric objects that you have observed, you would get more than one file with the data for those individual objects. It then proceeds with the comparison between the flow of the object under investigation and the flow of the other items measured in the field. Equation 1 allows us to obtain the values that are fitted by a linear function, that determines the best performance of our points of the curve of light. The sum in the denominator is used to obtain a theoretical star whose flux is equal to the sum of the fluxes of nearby individual stars. In this way you can build a chart, which you can do if precision is desired. It would show the highlights in the data, as in the brightness variations which would appear as positive or negative peaks in the light curve  (\cite{2014arXiv1409.2693S}). 

\subsection{Transiting Extra Solar Planets Inventory}

The Extra Solar Planets (ESP) census in the solar neighborhood started about 30 years ago, and while writing this paper, there are around 5600 extra solar planets known to man. Most of them are orbiting around either a G or K dwarf star, some may be going around slightly evolved stars. There are only a small number around M dwarfs, which is the Earth is that we are orbiting. It is important to remind us that this sample is not representative of the true ESP population in our galaxy, this is because the ESP survey is mainly focused on solar type stars. The resulting population of the ESP survey is composed of planets mainly with masses between a few Earth masses and 20 Jupiter masses, orbital periods ranging from 1 day to about 15 years, and the orbital eccentricity being between 0 and 0.996.

About one sixth of the ESP survey was surveyed from using the transit method. The transit method is the method of discovering exoplanets by using differential photometry to then be able to measure the occultation of a planet. This only really works the planets that are in the inner solar system.

Transiting planets can be mistaken as to being a possible eclipsing binary. With using the same methodologies used for studying these objects it is possible to determine the planets and its stars radii. The measured radii of the planets range from a few Earth radii to about 2 Jupiter radii. Some planets have a radii that is anomalously larger than the expected value from theoretical speculations, this could be due to tidal heating or due to the strong stellar irradiation which could inflate the planetary radius (\cite{2020ASPC..525...57H}).

\subsection{Duration of the Observations}

Transits of extra-solar planets have a typical duration of around 3 to 4 hours. This value highly depends on the orbital period, the planet, and its stars radii. To obtain a light-curve that is useful for a scientific analysis it is necessary to have a long time series before and after the transit. The Out-Of-Transit (\cite{2019ascl.soft04024P}) part of a light-curve allows us to estimate the value of depth of the transit and to correct for the presence of systematic (e.g. reference stars with colors very different from the target star). A practical rule for the observations of a transit is to obtain at least data from one hour before the transit up to one hour after the transit. The duration of a planetary transit (\cite{2005ApJ...627.1011T}) )) can be obtained with the Equation 3. This formula is accurate down to a few percentage points for the planets that may have very eccentric orbits ($e > 0.8$)

\begin{equation}
    D=\frac{2(R_{s}+R_{p})r_{t}}{\sqrt{G(M_{s}+M_{p})a(1-e^2)}}\sqrt{1-\frac{r_{t}^2\cos^2{i}}{(R_{s}+R_{p})^2}}
	\label{eq:D}
\end{equation}

Where $R_{s}$ is the star radius, $R_{p}$ is the planet radius, $M_{s}$ is the star mass, $M_{p}$ is the planet mass, $G$ is the gravitational constant, \textit{\textbf{a}} is the planetary orbital semi-major axis, \textit{\textbf{e}} is the planetary orbital eccentricity, \textit{\textbf{i}} is the planet inclination, and

\begin{equation}
    r_{t}=\frac{a(1-e^2)}{1+e\cos({\pi/2-\omega})}
	\label{eq:rt}
\end{equation}

\textbf{$\omega$} is the argument of periastron for the planet. Equation \ref{eq:rt} is calculating the depth of transit in flux units.

\begin{equation}
    \frac{dL}{L}=\biggl(\frac{R_{p}}{R_{s}}\biggr)^2
    \label{eq:flux}
\end{equation}

Where \textit{\textbf{L}} is luminosity. Equation \ref{eq:flux} determines the depth of transit in magnitudes. 

\begin{equation}
    \delta M = -2.5\log_{10}\biggl(1-\biggl(\frac{R_{p}}{R_{s}}\biggr)^2\biggr)
    \label{eq:delta}
\end{equation}

If the radius of the star is known (for example from the spectral classification), $R_{p}$ can be obtained from Equation \ref{eq:delta}; if the orbital period $P$ and the mass of the star $M_{\odot}$ are also known, the orbital semi-major axis, \textit{a} can be obtained from Kepler's third law (\cite{1989fnr..book...43D}), and therefore the duration of the transit can be obtained.

\begin{equation}
    \tau = \frac{P}{\pi}\biggl(\frac{R_{s}\cos \delta + r_{p}}{a}\biggr) \leq \frac{PR_{s}}{\pi a}
    \label{eq:tau}
\end{equation}

$\delta$ is the latitude of the transit over the stellar disk. The duration of the transit of Jupiter and the Earth for $a\cos i=0^{\circ}$ (equatorial transit) is $25h$ and $13h$ respectively (\cite{article}). From the previous equation we can obtain $\delta$, and therefore the inclination by using Equation \ref{eq:i}.

\begin{equation}
    \cos i = \frac{R_{s} \sin \delta}{a}
    \label{eq:i}
\end{equation}

\begin{figure}
\centering
\includegraphics[width=7cm]{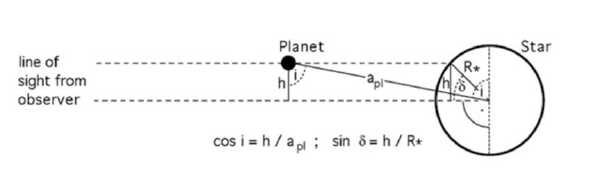}
\caption{Geometry of a planetary transit in which the relationship between the latitude of the transit on the star, $\delta$, and the inclination of the planetary orbit \textit{i} is shown. The impact parameter $h$ is the shortest projected distance between the planet and the center of the star.}\label{i}
\end{figure}

\subsection{Telescope Requirements}

The minimum telescope diameter for obtaining useful data from the observations is greater than $8cm=3.15inch$. Given that the photometry performed on bright stars will require out of focus images, the requirements on the optical quality of the mirror surfaces is moderate, with $\lambda/4$ being sufficient (\cite{2019arXiv190502790H}).

\subsection{Filters}

Images must be taken with the reddest filter that is available. This is to break down the problems of extinction in the field, the best choice would be a \textit{I filter}. At these wavelengths a back illuminated CCD chip could suffer the fringing effect (\cite{2012PASP..124..263H}), which is a difficult problem to deal with. For the homogeneity in light curves that is obtained, it is suggested to observe only with using \textit{R filters} with any kind of CCDs (back or front illuminated), using a \textit{R filter} we can combine light-curves more easily. It is not required that the filter is a photometric filter, but it is necessary that the filter cuts the blue wavelengths, due to the blue wavelengths being largely concerned by the scintillation effect (\cite{2015MNRAS.452.1707O}). 

If you don't have an \textit{R filter} you can use either a \textit{V filter} or an \textit{I filter}. The use of narrow band filters does not help in the observations of transits. The main purpose of choosing the previous types of filters noted is to obtain the largest number of photons. With narrow band filters the number of photons received by unit time is lower than with a wide band filter. The narrow band filters may also have the characteristics of being too sensitive to certain emission lines in the spectra of the stars and may not have certain spectra lines that would be needed for the target star. Comparison of fluxes obtained in these particular regime could be problematic because the measured value will be very sensitive to changes in the background and/or weather conditions.

\subsection{Ground-Based Observations: Light Curve Analysis}

\begin{figure}
	\includegraphics[width=\columnwidth]{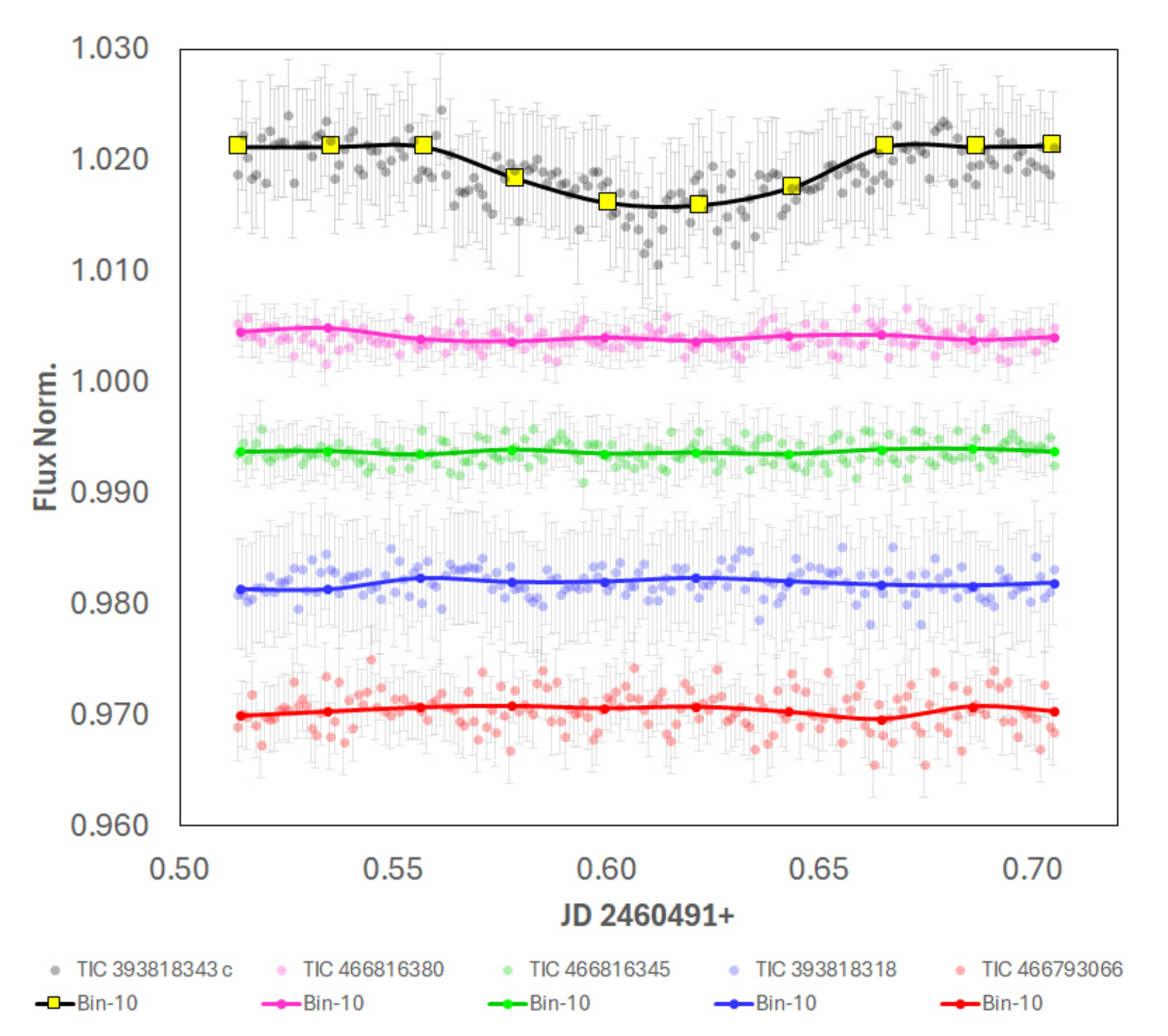}
    \caption{Differential photometry with comparisons between the target star and the reference stars (June 30, 2024 - \textit{ip Filter}).}
    \label{fig:20240630_ip}
\end{figure}

After the discovery of the planet TIC 393818343 b, our team scheduled off-transit observations to check for possible brightness variations due to the fact that TIC 393818343 b's system is double. The favourable weather allowed us to observe our system on the night of 30 June 2024, but only with the 0.40 m LCO telescope (Table \ref{tab:setup}).
The light curve obtained is shown in Figure \ref{fig:20240630_ip} together with the photometric flux of the comparison stars.

Being out of transit, a flat curve was expected. Instead, as can be seen in Figure \ref{fig:20240630_ip}, we detected a clear negative peak with magnitude $\delta m \sim 0.006$. Ingress, Mid-Transit, and the Output were clearly identified.  After analyzing the flux trend of the comparison stars, which did not show enough variation to suggest a false positive, and the fact that the event was complete, we were able to calculate some main parameters and in particular the period of the possible transit of a second planet belonging to the TIC 393818343 system. We then calculated the ephemeris of the possible transits to follow using the ExoFast tool (Section \ref{30th}, \ref{7th} and \ref{31th}).

The observation made with the defocus method provided a light curve with error ($\sigma \sim 1$) . In fact, by defocusing the star and optimising the exposure time, a very high degree of precision was achieved. Brightness variations on the order of thousandths of a magnitude were able to be seen. To reduce scintillation, we refer to Eqaution \ref{eq:scint} for calculating the approximate contribution of the scintillation to the photometric errors in terms of relative flux $dL/L$ (\cite{1998PASP..110..610D}):

\begin{equation}
    \sigma_{scint} = \frac{0.09d^{-2/3}A^{1.75}e^{-h/8000}}{\sqrt{2t_{exp}}}
    \label{eq:scint}
\end{equation}

 where \textit{d} is the telescope diameter in centimeters, \textit{A} is the air mass, \textit{h} is the height over sea level in meters, $t_{exp}$ is the exposure time in seconds. In Figure \ref{fig:scint_tab}  shows the expected scintillation for typical diameters and exposure times. To obtain the best possible lightcurve is necessary to minimize the contribution of scintillation. As the scintillation is dominant noise source for telescopes with diameters less than $40cm(15.75in)$, the best solution for decreasing this contribution is to have long exposure times. With long exposure times the \textit{high} frequency contribution of the scintillation will be averaged (Table in Figure \ref{fig:scint_tab}). For faint stars, $V > 10$, long exposure times are already necessary due to the low number of photons that reach the telescope. On the other side for very bright stars, obtaining long exposure time it is necessary to strongly defocus the star. In this case exposure time must be long enough to have a maximum value for the scintillation equal to 0.002 (\cite{10.1093/mnras/stad2964}).

 \begin{equation}
 A = \sec{z} - a_{1} \sec{(z-1)} - a_{2} \sec^2{(z-1)} - a_{3} \sec^3{(z-1)}
 \label{eq:airmass}
 \end{equation}

 where $\sec{z} = 1/\cos{z}$ and $z$ is the zenith distance, $a_{1}=0.0018167$, $a_{2}= 0.002875$, $a_{3} = 0.0008083$ (\cite{1989ApOpt..28.4735K}).

 \

 In our case, the transit depth of only 0.006 magnitudes is not easy to detect unless the observation setup and variables described are optimized.

 In light of the June 30th result, we then scheduled the observation, which, according to the calculated ephemeris, was to be made on July 7th. Naturally, knowing that the calculations were very approximate, we tried to use several observers. Unfortunately, LCO was unable to observe due to unfavourable weather, while MPC-C82 and MPC-L92 (Table \ref{tab:setup}) managed to obtain a series of images in the time interval corresponding to the ephemeris (Appendix Section \ref{exo}).

\begin{figure}
	\includegraphics[width=\columnwidth]{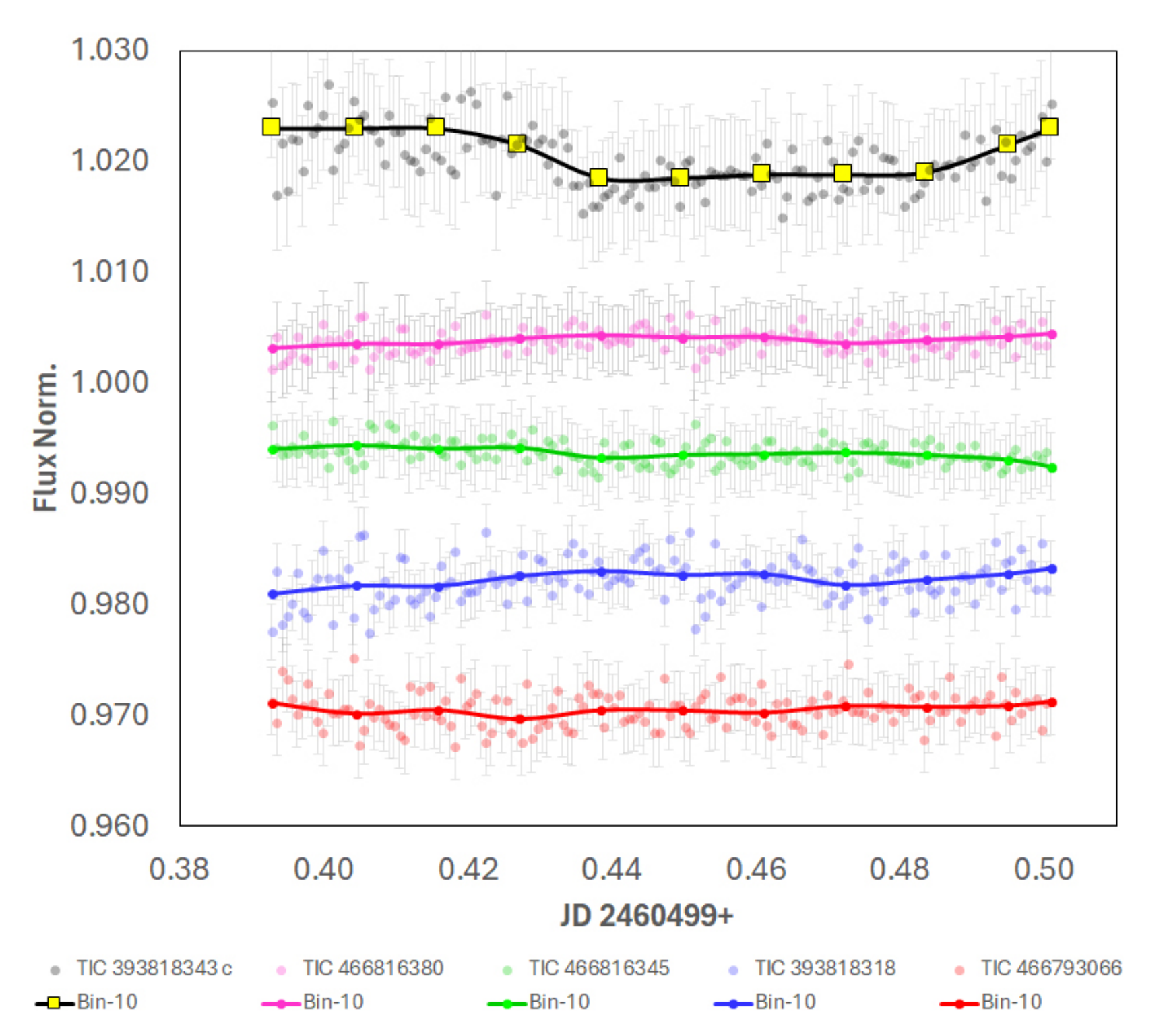}
    \caption{Differential photometry with comparisons between the target star and the reference stars (July 7, 2024 - \textit{Clear}).}
    \label{fig:20240707_Clear}
\end{figure}

\begin{figure}
	\includegraphics[width=\columnwidth]{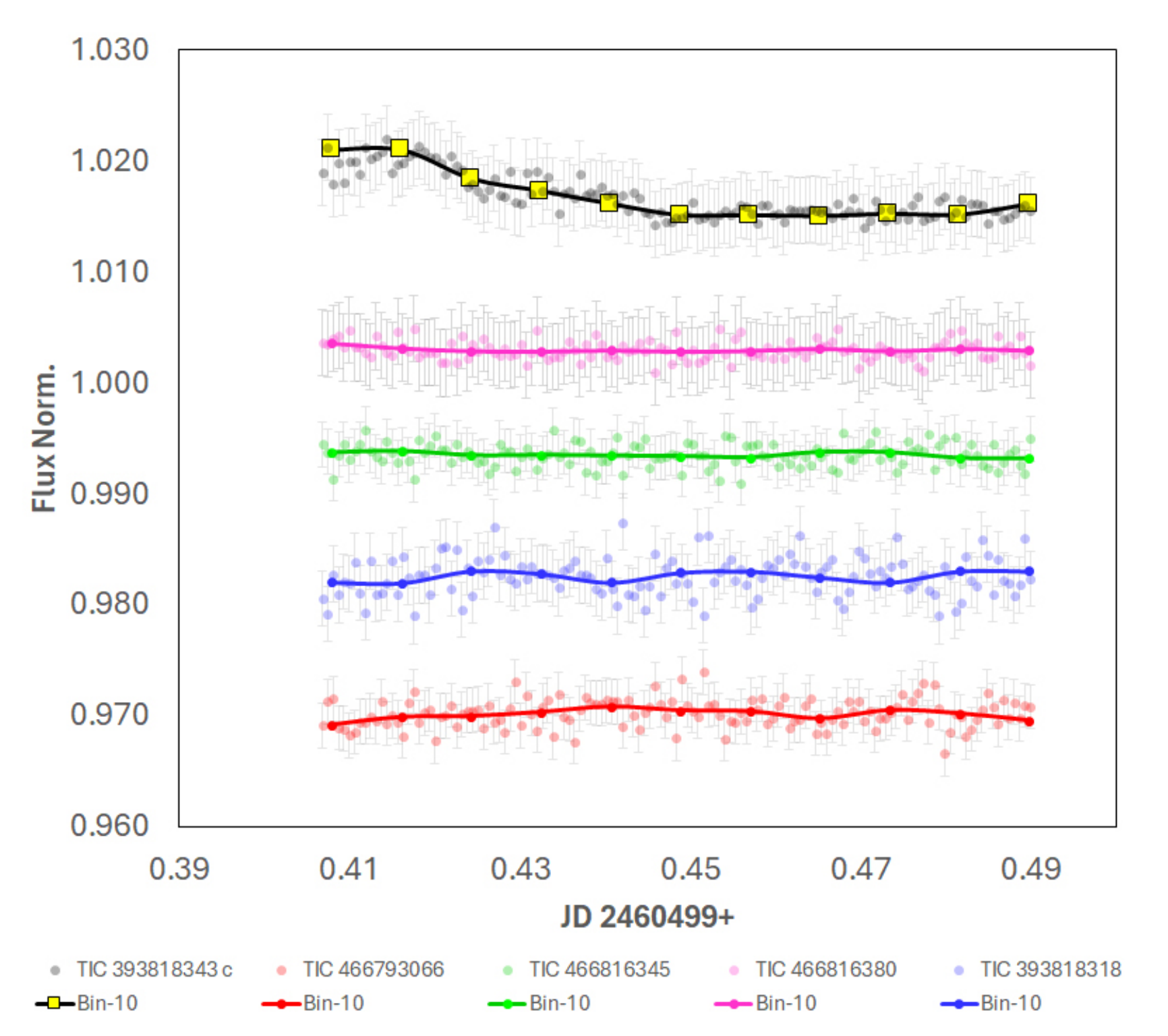}
    \caption{Differential photometry with comparisons between the target star and the reference stars (July 7, 2024 - \textit{Rc Filter}).}
    \label{fig:20240831_ip}
\end{figure}

\begin{figure}
	\includegraphics[width=\columnwidth]{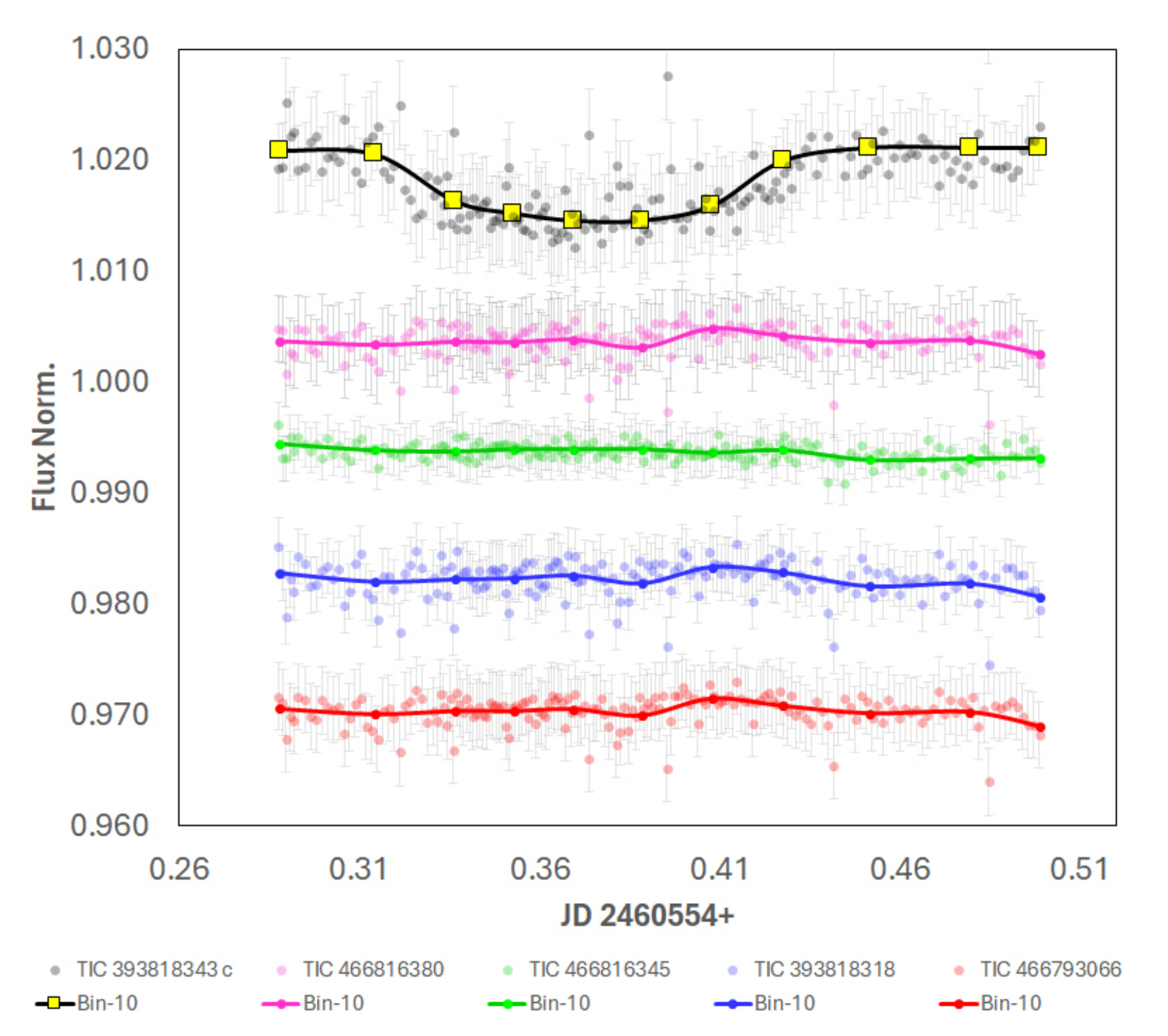}
    \caption{Differential photometry with comparisons between the target star and the reference stars (August 31, 2024 - \textit{Rc Filter}).}
    \label{fig:20240831_Rc}
\end{figure}

It is clearly evident that the July 7th observations are in perfect agreement with the June 30th observation in terms of both transit depth and duration. This outstanding result allowed us to redefine more precisely both the period and the ephemeris for the following observation and to see that both the photometric \textit{Rc filter} and the CMOS unfilter but with defocus were winning choices. Unfortunately, until August 31st the weather did not allow for any further observations.

For the August 31st event, all the team's observers were therefore alerted, so that at least one would be able to produce a light curve. Fortunately, MPC-C82 had the favourable weather to observe throughout the ephemeris interval.

As can be seen in Figure \ref{fig:20240831_ip}, the event was detected in its entirety with the same characteristics as the other events, thus unequivocally confirming the highly probable presence of another planet in the TIC 393818343 system. 

\begin{figure}
	\includegraphics[width=\columnwidth]{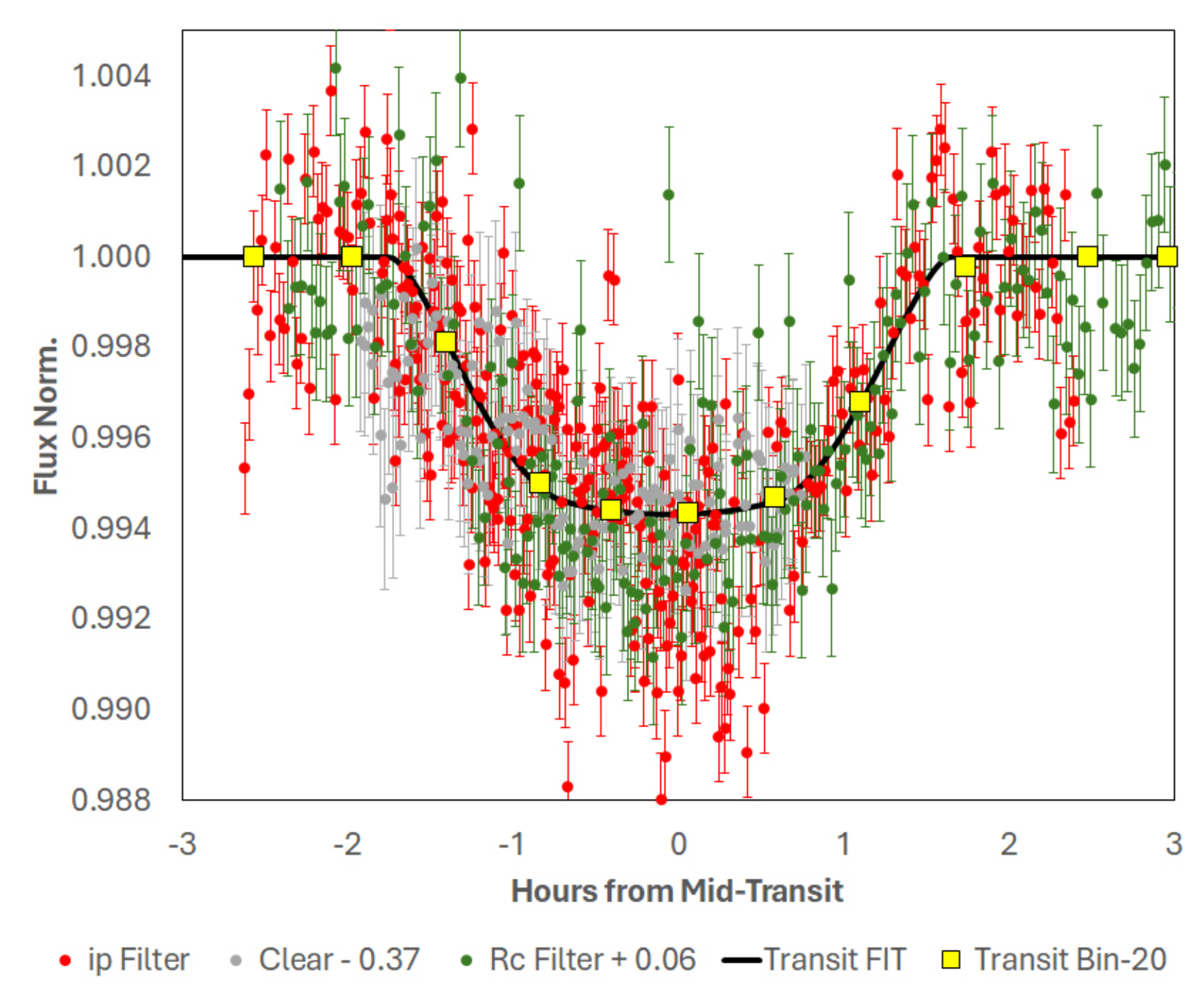}
    \caption{Observation of TIC 393818343 c planet transit using different photometric filters (Zero-Point on \textit{ip Filter)}, different telescopes and different locations (see Table \ref{tab:setup}).}
    \label{fig:CPF}
\end{figure}

\section{Synthetic Radial Velocity and Mass Investigation}

\subsection{Synthetic Radial Velocity}

Radial velocity measurements are necessary to classify Warm- Jupiters from the possibilities that might arise against the mass- radius problem, which may classify the planets as an eclipsing binary false positive. The probability of using TESS data (\cite{2015JATIS...1a4003R}), which we don’t comprise any of our data from, is shown at around $44\%$ (\cite{car2023}). The proof of the possibilities of false positives from the TESS data is a sign that ground-based observations are necessary. Radial velocity measurements have been used on a plethora of different papers to confirm an exoplanet. This method is shown from the previous confirmation of a TIC 393818343 planet (\cite{2024RNAAS...8...53C} and \cite{2024AJ....168...26S}), HD11964 b (\cite{2006ApJ...646..505B}), Wolf-327 b (\cite{2024A&A...684A..83M}), and many others.

Since radial velocity is not available to the team, we must run synthetic radial velocity possibilities. Synthetic radial velocity is a method used to calculate a possible mass from the radius using numerous Python packages, also includes a model, which must be used to estimate the mass for TIC 393818343 c. Using the synthetic radial velocity method, we used the public GitHub package \textbf{\textbf{Forecaster}} (\cite{2017ApJ...834...17C}). This model outputs four different measurements that could be possibly found with a selected planet to distinguish a planet from a false positive. These measurements range from a Terrestrial body all the way to a stellar- like possibility. It also includes the possibilities for a Neptunian or a Jovian-like planet, which allowed us to confirm that TIC 393818343 c is in those sections.

\subsection{Methodology}

The \textbf{Forecaster} model does have some variability in its code. This is to show a wide range of probabilities due to the inaccessibility of an equation for the mass-radius expectation. The code uses a Monte Carlo simulation to calculate the probabilities. There is over
1.4 billion probabilities of mass that could be calculated. Our team used a Monte Carlo simulation for about 500.000 possibilities. Even though it seems like an insignificant amount, it gives our team a good baseline to determine the outcome. With the use of the  \textbf{Forecaster} model and now knowing about the variability aspect, we can calculate the mass possibilities.

Before that, we must figure out the parameters that we will be using. The only data that we will be using is the radius of the planet. This data was derived from the June 30th, 2024, data. The data that we received from Ground-Based transit was $R_{p} = (8.683 \pm 0.198)R_{\oplus}$. There was a standard deviation of $\sigma \sim (0.707 \pm 0.155)$, this is for the radius which we have to use to have a more certain probability of the mass size on TIC 393818343 c. When the synthetic radius came back, we calculated the sigma confidence of $\sigma \sim (2.59 \pm 0.03)$ for this planet to not be a false positive (\cite{2018haex.bookE...4W}).

\subsection{Mass detection}

After receiving the model's thousands of calculations, they show a mass for TIC 393818343 c of $M_{p} = (113.763 \pm 41.999)M\oplus$ (Table \ref{tab:parameters_table}). From this information, we can show that it is most likely a Super-Neptune, and also a gas giant which is explained more in Section \ref{density}. We have learned from the model that there is a slight possibility of this planet not being a Super-Neptune. We will also explain why the planet isn’t the extreme possibilities, which will be explained later. As said previously a Super-Neptune is a planet that is larger than Neptune and has roughly a mass of $150M_{\oplus}$.

This planet is most likely not a terrestrial planet. This planet to be a possible terrestrial would be almost impossible, it has a lower-sigma mass of $M_{p} = (37.931 \pm 3.074)M\oplus$, which even from this measure shows a planet with a very low-density which wouldn’t make it a terrestrial planet. The lower-sigma mass shows characteristics of a planet that would still be significantly larger than a normal terrestrial planet.

From the transits, we can prove that this is not a stellar-like planet. This is shown by the transits not fluctuating. During a nearby eclipsing binary transit, the transit dip fluctuates (\cite{2011AJ....142..160S}). From our multiple observations, there were no fluctuations from the transit dips. During the investigation, it is considered that a sunspot could be causing the dips, which could happen if it was at a solar maximum. That has been proven false though, as sunspots cause irregularities in a light curve and not a clean light curve (\cite{2011ApJ...743...61S}).

From the information that we have retrieved we have determined that there is no possibility of a terrestrial-like planet or a stellar-like planet. Our final synopsis on this planet is that it must be a Super-Neptune.

\begin{figure}
	\includegraphics[width=\columnwidth]{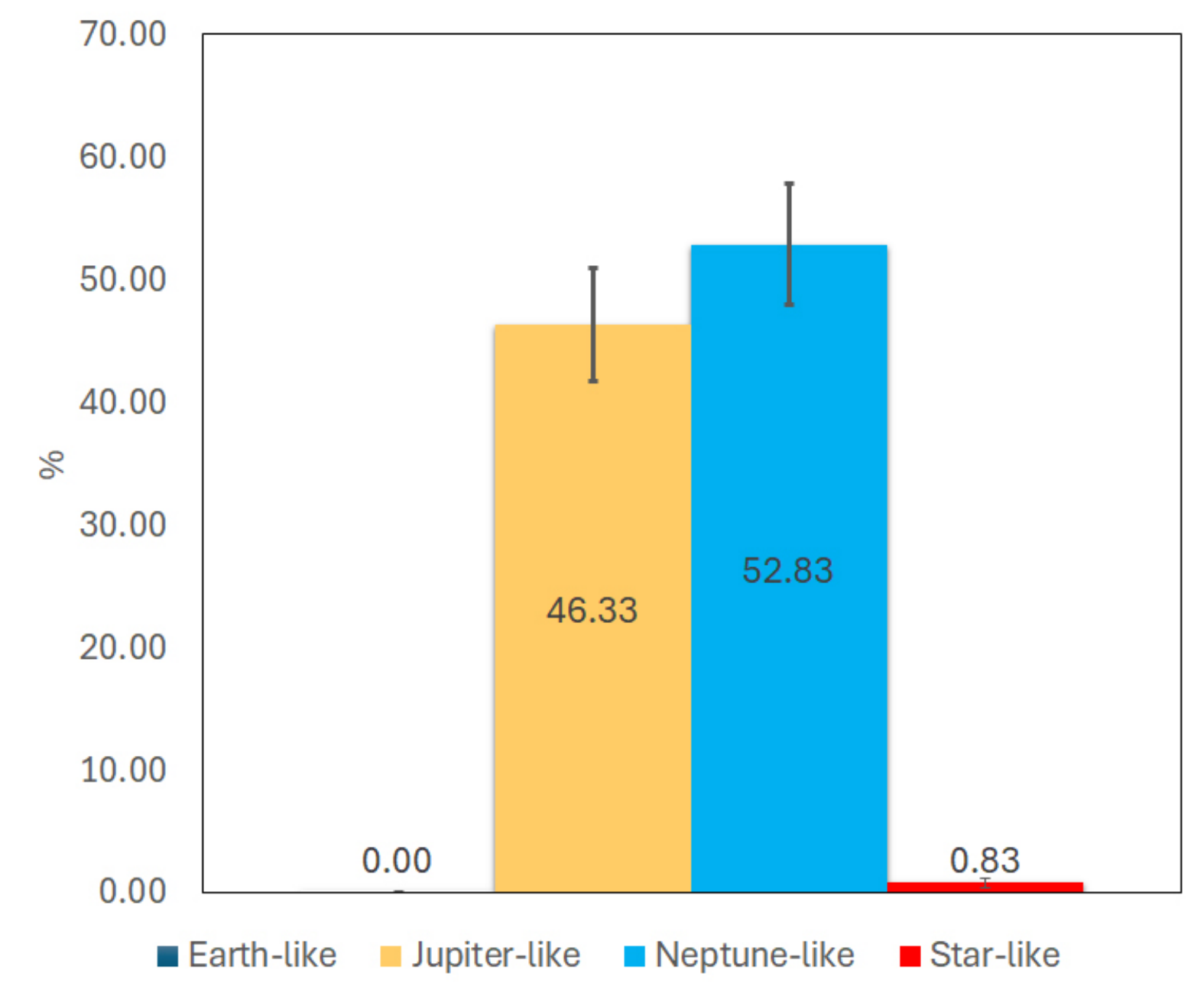}
    \caption{Probabilities with error margins on what the exoplanet mass could be from the synthetic radial velocity calculations.}
    \label{fig:torta}
\end{figure}

\begin{table}
	\centering
	\caption{Probabilities for each measurement to be the confirming measurement of mass for the exoplanet.}
	\label{tab:stat}
	\begin{tabular}{lccr} 
		\hline
		Earth-like & Jupiter-like & Neptune-like & Star-like\\
		\hline
		$0\%$ & $46.333\%$ & $52.833\%$ & $0.8\%$\\
		$\pm 0\%$ & $\pm 4.579\%$ & $\pm 4.926\%$ & $\pm 0.4\%$\\
		\hline
	\end{tabular}
\end{table}

\subsection{Density}\label{density}

Density is used as a measurement to guess the possibilities of the make up of the planet's composition. It also allows us to confirm the probabilities of what this planet could be. From the previous information from only the mass that it is a Super-Neptune.

The density of TIC 393818343 c is most likely as gas giant, with an average density of $(0.9585 \pm 0.3599) g/cm^{3}$. A gas giant is roughly between $0.7$ and $1.4$ $g/cm^{3}$, which is exactly where TIC 393818343 c falls. This information shows us that the planet across all data points
shows a gas giant that is a Neptunian-sized planet.

\section{Conclusions}

Through ground-based observations with several telescopes located in different locations around the world, the presence of a second planet orbiting the star TIC 393818343 with a period $P = (7.8458 \pm 0.0023) days$,was confirmed. Two complete and two partial transits were observed with photometry, allowing the estimation of ephemerides for subsequent observational follow-ups. It revealed a radius of $R_{p} = (0.792 \pm 0.018)R_{j}$ and an orbit with eccentricity $e = (0.389581 \pm 0.000015)$ using ExoFAST tool for best fitting. 

Light curves of transits in the \textit{ip}, \textit{Rc} and \textit{Clear} filters were produced, and the data were found to be perfectly compatible. The synthetic radial velocities estimated a mass $M_{p} = (0.357 \pm 0.132)M_{j}$ using the Forecast algorithm, and a density $\rho = (0.389581 \pm 0.000015)$.

TIC 393818343 c has been confirmed through ground-based
observations, rather than TESS data. That forces more questions to be asked with TESS reliability in detecting multiple exoplanets in the same system. In the context of multi-planetary systems, ruling out false positive scenarios is critical to accurately classifying exoplanets. This could either be from Near Eclipsing Binaries to some abnormalities with the star. The observations from ground-based telescopes have proved the presence of TIC 393818343 c. This is confirmed from simultaneous observations and through different models. More observations need to occur for the ephemeris of the planet to be fully solidified. TIC 393818343 c has a considerate background of information to uncover. This comes from the aspect of atmospheric spectroscopy (\cite{2017PASP..129c4401N}) to the true mass of the planet using radial velocity. Depending on what needs to be confirmed, TIC 393818343 c has been solidified, currently, as a Neptunian gas giant planet.

\section{Acknowledgment}

This research has made use of the Exoplanet Follow-up Observation Program (ExoFOP; DOI: 10.26134/ExoFOP5) website, which is operated by the California Institute of Technology, under contract with the National Aeronautics and Space Administration under the Exoplanet Exploration Program.




\bibliographystyle{mnras}
\bibliography{biblio} 







\onecolumn

\appendix

\section{Table of setup and locations}

\begin{table}
	\centering
	\caption{Observatories, locations and setup parameters used to detect TIC 393818343 c planet.}
	\label{tab:setup}
	\begin{tabular}{ lrrr } 
		\hline
		Parameter & Toni Scarmato's Observatory & Nastro Verde Observatory & LCO-Telescopes Observatory \\
   Type & MPC-L92 - Briatico (Italy) &  MPC-C82 - Sorrento (Italy) & MPC-Z21 - Teide (Spain) \\
		\hline
       Aperture & $250mm$ & $350mm$ & $350mm$ \\
       Focal Length & $1200$ $f/4.8$ & $3554$ $f/6.3$ & $1600$ $f/4.6$  \\
       Scale & $0.79$ $arcsec/pixel$ & $1.28$ $arcsec/pixel$ & $0.74$ $arcsec/pixel$ \\
       Optic & Newton & RC & Newton \\
       Type & Reflector & Ritchey-Chretien & Reflector \\
       Camera & ASI294MC PRO & SBIG ST-10 XME & QHY600M \\
       Filter & Clear (No Filter)& Rc & ip \\
       FOV & $54.8$’ $\times$ $37.1$’ & $16.3$’ $\times$ $16.3$’ & $1.9^{\circ}$ $\times$ $1.2^{\circ}$ \\
       FWHM & 300 $\Lambda$ & 200 $\Lambda$ & 1574 $\Lambda$  \\
       Area in Pixel array (BIN 1X1) & $4144px \times 2822px$ & $2184px \times 1472px$ & $9600px \times 6422px$ \\
       Pixel size & $4.63\mu \times 4.63\mu$& $6.8\mu \times 6.8\mu$ & $3.76\mu \times 3.76\mu$ \\
       Full well depth & $64Ke-$ & $77Ke-$ & $81Ke-$  \\
       Dark current & $<1e-/sec$ (at $-40^{\circ}C$)& $<1e-/sec$ (at $0^{\circ}C$) & $<1e-/sec$ (at $-20^{\circ}C$) \\
       Quantum efficiency & $75\%$& $86\%$ & $80\%$ \\
       A-D converter & $14bit$& $16bit$ & $16bit$ \\
       Readout Noise (RN) & $7.3 e-$& $8.8 e-$ & $3.7 e-$ \\
       Anti-blooming & Yes & No & Yes \\
       Cooling & Yes & Yes & Yes \\
       Type & CMOS & CCD & CMOS \\
       Sensor size & $19.2mm \times 13.0mm$ & $14.9mm \times 10.0mm$  & $35mm \times 35mm$\\
		\hline
	\end{tabular}
\end{table}

\section{Table of reference stars}

\begin{table}
	\centering
	\caption{The reference stars used for differential photometry with indication of magnitude in the filters adopted \citep{1996AJ....111.1748F}.}
	\label{tab:ref_stars}
	\begin{tabular}{lccccr}
		\hline
		Ref & Star & Coordinates & Magnitude & Magnitude & Magnitude\\
   N$^\circ$ & name & J2000 & (ip Filter) & (Clear) & (Rc filter)\\
		\hline
	      Ref1 & TIC 466816380 & $20:41:46.08, +03:36:56.91$ & 9.31 & 9.68 & 9.25 \\
            Ref2 & TIC 466816345 & $20:41:26.94 ,+03:39:15.29$ & 12.61 & 12.98 & 12.55 \\
            Ref3 & TIC 393818318 & $20:41:19.28 ,+03:39:55.17$ & 12.99 & 13.36 & 12.93 \\
            Ref4 & TIC 466793066 & $20:40:45.92 ,+03:34:56.67$ & 12.06 & 12.43 & 12.01\\
		\hline
	\end{tabular}
\end{table}

\newpage

\section{Scintillation}

\begin{figure}
\centering
	\includegraphics[width=11cm]{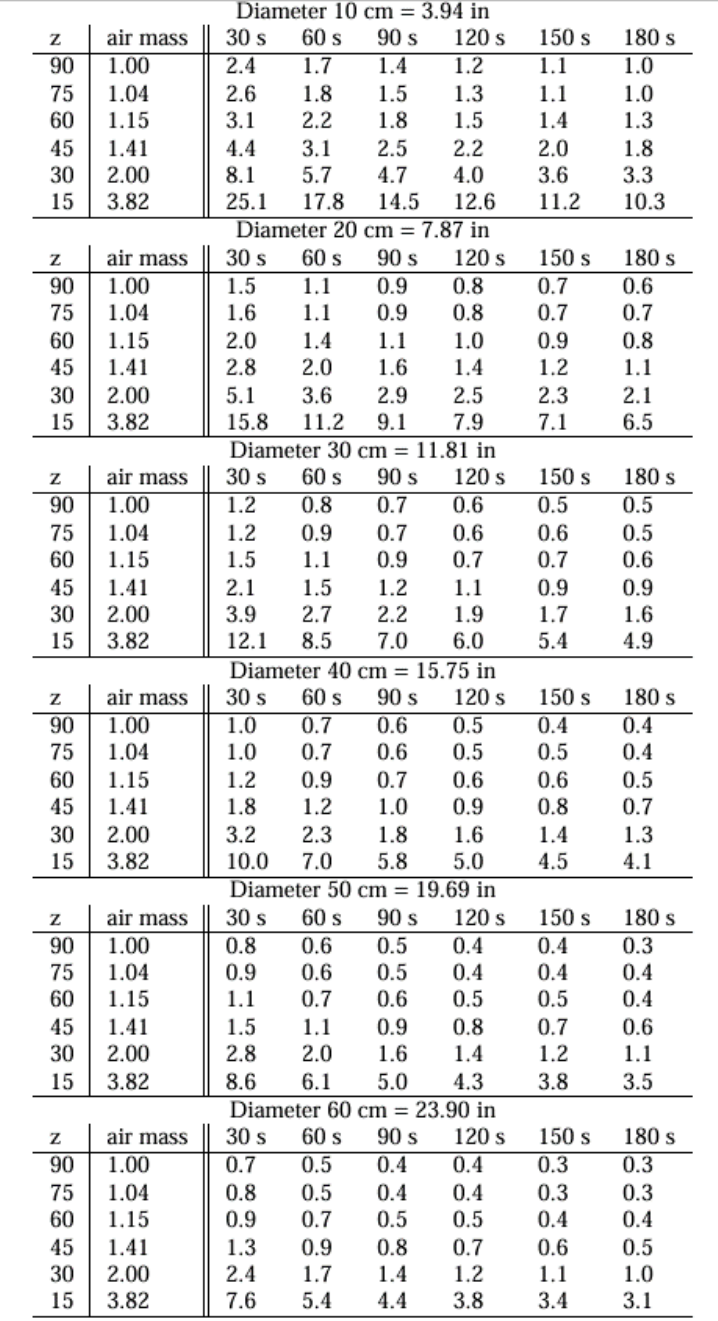}
    \caption{Scintillation values in $10^{-3}$ units of relative flux $dL/L$ for typical exposure time and some telescope aperture. Values for intermediate exposure times and aperture could be obtained by interpolation between adjacent values. Scintillation was calculated for an altitude of $300m(984.25ft)$ over the sea level. Lower altitudes provide greater values for scintillations, higher altitudes provide low scintillation values. Between $0$ and $1000m$($0$ and $3281ft$), the scintillation change only of 0.1mmag. Only over $2000m(6562ft)$, the scintillation start to decrease significantly.}
    \label{fig:scint_tab}
\end{figure}

\section{ExoFAST Outputs}\label{exo}

\subsection{30 June 2024 observation}\label{30th}

\begin{verbatim}
Transit fit:
Chi^2/dof = 641.02050
Scaling errors by 25.335145
RMS of residuals = 0.0025011656

Combined fit:
Chi^2 of Transit data = 426.10756 (516 data points)
Chi^2 of Priors = 11.315043 (10 priors)
Chi^2/dof = 0.86996945


Stellar Parameters:
          M_{*}                             Mass (\msun)       1.105199
          R_{*}                           Radius (\rsun)       1.090575
          L_{*}                       Luminosity (\lsun)       1.162508
         \rho_*                            Density (cgs)       1.202572
      \log(g_*)                    Surface gravity (cgs)       4.406247
          \teff                Effective temperature (K)    5743.436683
           \feh                               Metalicity       0.316355
Planetary Parameters:
              e                             Eccentricity       0.393425
       \omega_*         Argument of periastron (degrees)    -101.681760
              P                            Period (days)       7.826248
              a                     Semi-major axis (AU)       0.079741
          R_{P}                             Radius (\rj)       0.828492
         T_{eq}              Equilibrium Temperature (K)    1024.067640
          \fave                 Incident flux (\fluxcgs)       0.214902
RV Parameters:
  e\cos\omega_*                                               -0.079659
  e\sin\omega_*                                               -0.385276
          T_{P}             Time of periastron (\bjdtdb) 2460495.004200
Primary Transit Parameters:
            T_C                Time of transit (\bjdtdb) 2460491.622666
    R_{P}/R_{*}        Radius of planet in stellar radii       0.078090
        a/R_{*}         Semi-major axis in stellar radii      15.727381
            u_1              linear limb-darkening coeff       0.866639
            u_2           quadratic limb-darkening coeff      -0.007987
              i                    Inclination (degrees)      87.602314
              b                         Impact Parameter       0.904664
         \delta                            Transit depth       0.006098
       T_{FWHM}                     FWHM duration (days)       0.090575
           \tau           Ingress/egress duration (days)       0.048492
         T_{14}                    Total duration (days)       0.139067
          P_{T}        A priori non-grazing transit prob       0.042633
        P_{T,G}                    A priori transit prob       0.049855
            F_0                            Baseline flux       1.000264
Secondary Eclipse Parameters:
          T_{S}                Time of eclipse (\bjdtdb) 2460495.106234
          b_{S}                         Impact parameter       0.401449
     T_{S,FWHM}                     FWHM duration (days)       0.096352
         \tau_S           Ingress/egress duration (days)       0.008992
       T_{S,14}                    Total duration (days)       0.105344
          P_{S}        A priori non-grazing eclipse prob       0.096073
        P_{S,G}                    A priori eclipse prob       0.112348

Errors from Carter et al., 2008 (eqs 19 & 23):
\sigma_{T,C}    ~ 0.0012541586
\sigma_{\tau}   ~ 0.0043445327
\sigma_{T,FWHM} ~ 0.0025083171
\sigma_{\depth} ~ 0.00016319801

NOTE: depth used here (0.0060979763) is not delta
      if the transit is grazing
\end{verbatim}

\subsection{7 July 2024 observation}\label{7th}

\begin{verbatim}
Transit fit:
Chi^2/dof = 539.52405
Scaling errors by 23.240091
RMS of residuals = 0.0023010616

Combined fit:
Chi^2 of Transit data = 631.29737 (635 data points)
Chi^2 of Priors = 4.7948863 (10 priors)
Chi^2/dof = 1.0229077


Stellar Parameters:
          M_{*}                             Mass (\msun)       1.114585
          R_{*}                           Radius (\rsun)       1.100136
          L_{*}                       Luminosity (\lsun)       1.185147
         \rho_*                            Density (cgs)       1.181438
      \log(g_*)                    Surface gravity (cgs)       4.402338
          \teff                Effective temperature (K)    5746.063561
           \feh                               Metalicity       0.318880
Planetary Parameters:
              e                             Eccentricity       0.389595
       \omega_*         Argument of periastron (degrees)    -107.818554
              P                            Period (days)       7.825960
              a                     Semi-major axis (AU)       0.079964
          R_{P}                             Radius (\rj)       0.836473
         T_{eq}              Equilibrium Temperature (K)    1027.580584
          \fave                 Incident flux (\fluxcgs)       0.218474
RV Parameters:
  e\cos\omega_*                                               -0.119218
  e\sin\omega_*                                               -0.370906
          T_{P}             Time of periastron (\bjdtdb) 2460494.740115
Primary Transit Parameters:
            T_C                Time of transit (\bjdtdb) 2460491.623091
    R_{P}/R_{*}        Radius of planet in stellar radii       0.078157
        a/R_{*}         Semi-major axis in stellar radii      15.634323
            u_1              linear limb-darkening coeff       0.866637
            u_2           quadratic limb-darkening coeff      -0.007982
              i                    Inclination (degrees)      87.572083
              b                         Impact Parameter       0.893000
         \delta                            Transit depth       0.006108
       T_{FWHM}                     FWHM duration (days)       0.097251
           \tau           Ingress/egress duration (days)       0.043832
         T_{14}                    Total duration (days)       0.141083
          P_{T}        A priori non-grazing transit prob       0.043731
        P_{T,G}                    A priori transit prob       0.051146
            F_0                            Baseline flux       1.000490
Secondary Eclipse Parameters:
          T_{S}                Time of eclipse (\bjdtdb) 2460494.898249
          b_{S}                         Impact parameter       0.409788
     T_{S,FWHM}                     FWHM duration (days)       0.097713
         \tau_S           Ingress/egress duration (days)       0.009202
       T_{S,14}                    Total duration (days)       0.106916
          P_{S}        A priori non-grazing eclipse prob       0.095297
        P_{S,G}                    A priori eclipse prob       0.111456

Errors from Carter et al., 2008 (eqs 19 & 23):
\sigma_{T,C}    ~ 0.00097131581
\sigma_{\tau}   ~ 0.0033647367
\sigma_{T,FWHM} ~ 0.0019426316
\sigma_{\depth} ~ 0.00012851799

NOTE: depth used here (0.0061084626) is not delta
      if the transit is grazing
\end{verbatim}

\subsection{31 August 2024 observation}\label{31th}

\begin{verbatim}
Transit fit:
Chi^2/dof = 465.59797
Scaling errors by 21.619145
RMS of residuals = 0.0083856296

Combined fit:
Chi^2 of Transit data = 173.75625 (185 data points)
Chi^2 of Priors = 0.10608419 (13 priors)
Chi^2/dof = 1.0117797


Stellar Parameters:
          M_{*}                             Mass (\msun)       1.100317
          R_{*}                           Radius (\rsun)       1.094962
          L_{*}                       Luminosity (\lsun)       1.175103
         \rho_*                            Density (cgs)       1.182928
      \log(g_*)                    Surface gravity (cgs)       4.400837
          \teff                Effective temperature (K)    5747.383700
           \feh                               Metalicity       0.320087
Planetary Parameters:
              e                             Eccentricity       0.389566
       \omega_*         Argument of periastron (degrees)    -107.819234
              P                            Period (days)       7.843432
              a                     Semi-major axis (AU)       0.079740
          R_{P}                             Radius (\rj)       0.832657
         T_{eq}              Equilibrium Temperature (K)    1026.837242
          \fave                 Incident flux (\fluxcgs)       0.217847
RV Parameters:
  e\cos\omega_*                                               -0.119213
  e\sin\omega_*                                               -0.370877
          T_{P}             Time of periastron (\bjdtdb) 2460557.494543
Primary Transit Parameters:
            T_C                Time of transit (\bjdtdb) 2460554.370547
    R_{P}/R_{*}        Radius of planet in stellar radii       0.078168
        a/R_{*}         Semi-major axis in stellar radii      15.664162
            u_1              linear limb-darkening coeff       0.866633
            u_2           quadratic limb-darkening coeff      -0.007972
              i                    Inclination (degrees)      87.650519
              b                         Impact Parameter       0.865800
         \delta                            Transit depth       0.006110
       T_{FWHM}                     FWHM duration (days)       0.112002
           \tau           Ingress/egress duration (days)       0.038088
         T_{14}                    Total duration (days)       0.150089
          P_{T}        A priori non-grazing transit prob       0.043648
        P_{T,G}                    A priori transit prob       0.051050
            F_0                            Baseline flux       1.001875
Secondary Eclipse Parameters:
          T_{S}                Time of eclipse (\bjdtdb) 2460557.653050
          b_{S}                         Impact parameter       0.397332
     T_{S,FWHM}                     FWHM duration (days)       0.098338
         \tau_S           Ingress/egress duration (days)       0.009151
       T_{S,14}                    Total duration (days)       0.107489
          P_{S}        A priori non-grazing eclipse prob       0.095110
        P_{S,G}                    A priori eclipse prob       0.111240

Errors from Carter et al., 2008 (eqs 19 & 23):
\sigma_{T,C}    ~ 0.0068898612
\sigma_{\tau}   ~ 0.023867179
\sigma_{T,FWHM} ~ 0.013779722
\sigma_{\depth} ~ 0.00091154322

NOTE: depth used here (0.0061101978) is not delta
      if the transit is grazing
\end{verbatim}

\section{Table of parameters}

\begin{table}
	\centering
	\caption{Parameters table of TIC 393818343 c identified using the ExoFast tool.}
	\label{tab:parameters_table}
	\begin{tabular}{lcr} 
		\hline
		Parameter & Units & Values\\
		\hline
		$P$ & Period (days) & $7.8458 \pm 0.0023$\\
		$R_{p}$ & Radius($R_{J}$) & $0.792 \pm 0.018$\\
		$M_{p}$ & Mass($M_{J}$) & $0.357 \pm 0.132$\\
            $\rho$ & Density($g/cm^3$) & $0.9585 \pm 0.3599$\\
            $T_{c}$ & Time of transit ($BJD_{TDB}$) & $2460554.3703 \pm 0.0003$\\
            $a$ & Semi-major axis (AU) & $0.079 \pm 0.011$\\
            $i$ & Inclination (degrees) & $87.6503 \pm 0.0003$\\
            $e$ & Eccentricity & $0.389581 \pm 0.000015$\\
            $\omega_{*}$ & Argument of periastron (degrees) & $-107.8189 \pm 0.0003$\\
            $T_{eq}$ & Equilibrium Temperature (K) & $1027.2089 \pm 0.3717$\\
            $R_{p}/R_{*}$ & Radius of planet in stellar radii & $0.078163 \pm 0.000006$\\
            $a/R_{*}$ & Semi-major axis in stellar radii & $15.649 \pm 0.015$\\
            $\delta$ & Transit depth & $0.006109 \pm 0.000001$\\
            $\tau$ & Ingress/egress duration (days) & $0.041 \pm 0.003$\\
            $T_{14}$ & Total duration (days) & $0.1456 \pm 0.0045$\\
            $b$ & Impact Parameter & $0.8765 \pm 0.0026$\\
            $b_{S}$ & Eclipse Impact Parameter & $0.3972 \pm 0.0002$\\
            $\tau_S$ & Ingress/egress Eclipse duration (days) & $0.009 \pm 0.001$\\
            $T_{S,14}$ & Total Eclipse duration (days) & $0.195 \pm 0.186$\\
            $T_{P}$ & Time of periastron ($BJD_{TDB}$) & $2460557.4943 \pm 0.0003$\\
            $T_{S}$ & Time of eclipse ($BJD_{TDB}$) & $2460556.0118 \pm 1.6412$\\
            $e\cos\omega_*$ &  & $-0.1192 \pm 0.0001$\\
            $e\sin\omega_*$ &  & $-0.3704 \pm 0.0004$\\
            $u_1$ & linear limb-darkening coeff & $0.866 \pm 0.003$\\
            $u_2$ & quadratic limb-darkening coeff & $-0.00799 \pm 0.00001$\\
		\hline
	\end{tabular}
\end{table}


\bsp	
\label{lastpage}
\end{document}